\documentclass{epl}

\usepackage{psfig}

\newcommand{\be}{\begin{equation}}
\newcommand{\ee}{\end{equation}}
\newcommand{\alt}{$ \raisebox{-1.mm}{$\stackrel{<}{\scriptstyle{\sim}}$} $}
\newcommand{\agt}{\stackrel{>}{\scriptstyle{\sim}}}

\title{
Fractals versus halos: asymptotic scaling without 
fractal properties
}
\shorttitle{Fractals vs halos....}
\author{
M. Bottaccio\inst{1,3} \and M.Montuori\inst{1,3} \and L. Pietronero\inst{1,2,3}}
\institute{
\inst{1} Center for Statistical Mechanics and Complexity -
Dep. Physics - University ``La Sapienza''- P.le A. Moro, 2  00185 - Roma, Italy\\
\inst{2} CNR  Istituto di Acustica ``O. M. Corbino'' 
- V. del Fosso del Cavaliere, 100  00133 - Roma, Italy\\
\inst{3} Centro Studi e Ricerche e Museo della Fisica ``E.Fermi'', - Compendio 
Viminale, - Roma, Italy\\
}

\begin{document}
\maketitle

\date{\today}

\begin{abstract}
  Precise analyses of the statistical and scaling properties of galaxy
distribution are essential to elucidate the large scale structure 
of the universe. Given the ongoing debate on its statistical features,
the development of statistical tools 
 permitting to  discriminate accurately different spatial patterns
are highly desiderable.

This is specially the case when non-fractal distributions have power-law two point
correlation functions, which are usually signatures of fractal properties.
Here we review some possible methods used in the litterature and 
introduce a new variable called "scaling gradient".
This tool and the conditional variance are shown to be effective in providing 
an unambiguous way for such a 
distinction. 
Their application is expected to be 
of outmost importance in the analysis of upcoming galaxy-catalogues.
\end{abstract}
 

Understanding the statistical properties of the spatial distribution of matter in
the universe is a fundamental issue in cosmology and astrophysics.
 It provides an important tool to test  the features of cosmological models
 and it is intimately related to the nature of the matter
distribution and the 
dynamical processes which have  shaped the present universe.
 During the past twenty years observations 
have revealed a  hierarchy of structures (termed large scale structure, LSS):
galaxies are grouped in clusters, which in turn appear
to form larger associations, the superclusters, separated by
 wide nearly empty regions 
	
These structures have been characterized mainly through their correlation
properties, in particular by the two-point correlation function.
Such studies have found the presence of power law two-point correlations
in  a wide range of scales. 
Many authors have interpreted such behavior as the signature
of a fractal (or even multifractal) \cite{de:vauc:70,peebles:76,brot:75,col:piet:88}.
However, in many cases, the conceptual and practical implications of a fractal distribution 
have not been really considered~\cite{piet:87,col:piet:88}.

In fact, one of the implications of fractal correlations is that one 
cannot define the eventual crossover length from the usual correlation function.
This point has generated a large debate in the field~\cite{piet:87,colem:92,physrep,peebles:89,guzzo:97,wu:99,mart:99}.
In tab.~\ref{tab1} we present a comprehensive summary of the properties of galaxy correlations, as obtained
with various methods.
The main results are
the value of the fractal dimension $D$ and the eventual crossover length to a homogeneous distribution ($D=3$).
The estimation of such a scale varies from $10$ to $300$ Mpc $h^{-1}$ ($h$ is a constant $\approx 0.7$). 

In Sylos Labini et al.~\cite{physrep} it has been shown 
that galaxy  correlations  from different samples measured with more general statistical tools
are  consistent with each other and with a fractal dimension   $D\approx 2$, without a clear detection of any
 crossover to a homogeneous distribution
\begin{table}
\caption{\label{tab1}  Estimates  
of the galaxy distribution fractal dimension $D$ and of
the range (in Mpc $h^{-1}$) over which it extends
(if indicated in the corresponding paper). Note that the reported 
values of $D$ are obtained by different 
methods of measure; for this reason 
we choose the generic denomination 
of {\it fractal dimension} $D$. 
 * in fact, a 
 multifractal with dimension $D_2=1.3$ and $D_o=2$;  $\dag$   due to planes, rather than fractal;
 $\ddag$ homogeneity not evident
in the samples analysed; $\S$     specific galaxy samples.}
\begin{center}
\begin{tabular}{l|c|l}
Author & D & Range\\
\hline\hline
Mandelbrot (1975)~\cite{brot:75} &  1.3 & ? \\
\hline
Carpenter (1986)~\cite{carp:86}   &   2 $\rightarrow$ 2.8   & ?\\
\hline
Deng et al. (1988)~\cite{deng:88} &   2.0  & ? \\
\hline
Coleman et al. (1988)~\cite{col:piet:88} & 1.4 $\simeq$ 1.5  & 
$r\alt 28$  \\
\hline
Peebles (1989)~\cite{peebles:89}   &  1.23 & $r \alt 15$   \\ 
\hline
Martinez et al.(1990)~\cite{mart:90} &  $1.3 {\mbox{\large *}}$ &  $1 \alt r \alt 5$ \\
\hline
Luo \& Schramm (1992)~\cite{luo:92} & 1.2 & $10 \alt r \alt 100$ \\
\hline
Provenzale (1994)~\cite{provenz:94} & 1.2 & $r \alt 4 $ \\
                  & 2 \mbox{\normalsize \dag} & $4\alt r\alt25$  \\
\hline 
Guzzo (1997)~\cite{guzzo:97} & 1.2 & $r  \alt$ 3.5  \\
             &  2 - 2.3 & $3.5 \alt r \alt$ 20 -30  \\
\hline
Sylos Labini et al. (1998)~\cite{physrep} & 2 &    \ddag \\
\hline
Scaramella et al. (1998)~\cite{scaramella:98} &  $< 3$ &  $r \alt 300$   \\
\hline
Wu et al. (1999)~\cite{wu:99} &  1.2 - 2.2 &  $r \alt 10$   \\
                 &  tends to 3 & $10 \alt r \alt 100$   \\
\hline
Martinez (1999)~\cite{mart:99} &  2 & $r \alt 15 $   \\
                & 3 & $r \agt 30 $  \\
\hline
Pan \& Coles (2002)~\cite{pan:02} &  2.16 (PSCZ) $\S$ &   $r \alt 10$  \\
                    &  1.8 (Cfa2) $\S$ &  $r \alt 40$  \\
\end{tabular}
\end{center}
\end{table}
The detection of fractal properties in LSS raised the issue of their origin.
Many authors have claimed that fractal structures are naturally formed in 
cosmological N-body simulations (e.g.~\cite{valdarnini:92}) 
driven essentially just by gravitational interactions.

An alternative, very popular model which also tries to explain the 
power-law  correlations 
is the halo  model~\cite{peebles:74}.
This model takes also inspiration from the analysis
of  N-body simulations, where
small scale structures look like compact, almost spherical, clusters (halos), 
with little inner substructure (but see e.g.~\cite{moore:99}) 
rather than fractal. 
In this model,  two-point power-law correlations up to the halo size
are due 
to particles belonging to the same halo. The crucial point is that
 some kind of non-fractal cluster density profiles
can give  power law two-point correlations, like in a fractal distribution.

In this model, however, one does not expect to see 
a single power law from scales smaller
to  scales larger  than the halo size (few Mpc)
(tab.~\ref{tab1},~\cite{bahcall:88,physrep}).
The detection of a different behaviour of correlations in the two regimes
has actually been claimed in~\cite{astroph}.

There is a essential difference between this view where 
correlations are due to structures with a regular density profile
 and the fractal one.  Although such difference has been noted by some 
authors~\cite{murante:97}, there has been little attempt 
to discriminate in a quantitative
way which picture actually corresponds to the observed distribution, both for the galaxy data 
and N-body simulations.
In this letter we clarify this basic problem
from a conceptual and practical point of view.

In particular,  we show that specific statistical tools related to the three-point 
correlation analysis can be usefully applied to discriminate between the various scenarios. 
Moreover, we define a new concept (``the scaling gradient'') which 
appears particularly suitable in this respect.
The application of these methods to new, large catalogs will presumably
resolve the issue of the true statistical properties of the galaxy distribution.

We start by
considering
 the simple 
example of a halo characterized by a single power law density, firstly
 explored in 3d by~\cite{peebles:74,mcclell:77}.
Since then, there has been a large number of studies on the halo properties (for a review, see~\cite{cooray:02} 
and references therein). 
Actually, N-body simulations show halos with density profiles which can be approximated by a power-law 
only in a range of scales~\cite{navarro}. However, the profile we investigate here retains 
the relevant statistical features of realistic halos.

Assume a continuous density distribution in $d$ dimensions decaying   from
 its center as:
\begin{equation}
\rho(r) = Ar^{-\beta}
\label{eq1}
\end{equation}    
with $0<\beta<d$.

For simplicity in the following the formulas refer to systems of unit
size.
 Clearly, such a system is not a fractal:  there is only
one density singularity,  at the origin, and the distribution is 
analytical everywhere else.
Its
 density-density correlation (or conditional average density \cite{colem:92}) can be
estimated analytically:
\be
\Gamma^*(r)=\frac{1}{\bar \rho V(r)}\int_{V(r)} 
d{\bf r'}d^ds \langle\rho({\bf r'})\rho({\bf r' +s})\rangle=
 \frac{A}{d-\beta}\left\{ \frac{(d-\beta)^2}{d-2\beta}
+r^{d-2\beta}\left[d-\frac{(d-\beta)^2}{d-2\beta}\right]\right\}.
\label{eq4}
\end{equation}
where $\rho({\bf r'})$ is the density in ${\bf r'}$, ${\bar \rho}$ 
is the average density, the average $\langle\ldots\rangle$ is performed
over the angles between ${\bf r}$ and ${\bf r'}$ and over ${\bf r'}$, and 
$V(r)$ is the volume of a sphere of radius $r$.

Eq.~(\ref{eq4}) shows that for  $\beta < d/2$  
the first term in curly brackets dominates;
 therefore the average conditional density 
is constant, as in a homogeneous density
field.
\begin{figure}[h]
\centerline{
        \psfig{file=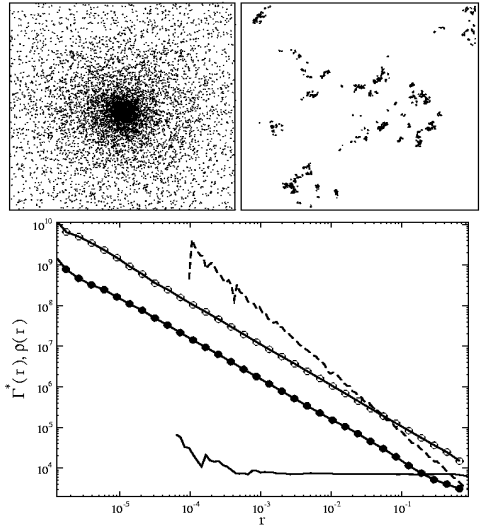,width=6.5cm,height=7.1cm}
}
\caption{
Top: left, $2d$ halo with $10000$ points
 and density given by eq.~\ref{eq1}, with $\beta=1.5$;
right, a fractal distribution with fractal dimension
$D=1$ generated by a Levy-flight algorithm ($8000$
points).
Bottom: Average conditional density for the set in the top left frame 
({\em empty
circles})
and the Levy flight fractal ({\em filled circles}). 
It is apparent that the two distributions
have the same scaling in $\Gamma^*(r)$.
The density profile of the halo with $\beta=1.5$ ({\em dashed line}) 
is steeper than the corresponding
   $\Gamma^*(r)$.
The $\Gamma^*(r)$ for a halo with $\beta=0.5$ is shown 
by the {\em solid line}.
}
\label{fig4}
\end{figure}
For $\beta > d/2$, instead, the second term dominates
and the average conditional density is a  power law
with exponent $d-2\beta$ {\it at any scale}. This behavior
appears therefore identical to the one of
 a fractal sample with dimension $D= 2d-2\beta$.
In fig.~\ref{fig4} we show that a halo and a fractal
can have precisely the same scaling in $\Gamma^*(r)$, even though they  are completely 
different systems.

In the light of this result, however, there has been little effort 
to clarify the difference between the two possibilities in the analysis
of LSS data and in N-body computer simulations

In principle, a distinction between different sets of points with the same 
two point correlation properties could be obtained using
box counting methods~\cite{feder}.
For the system described by eq.~(\ref{eq1}), we have:
\begin{equation}
\chi(q) = \lim_{l\rightarrow0} \sum 
\mu_i^q = 
 \lim_{l\rightarrow0} 
(B_1(\beta,q) l^{q(d-\beta)}+ B_2(\beta,q) l^{d(q-1)}),
\label{eq6}
\end{equation}
where $l$ is the box size, $\mu_i$ is the mass inside
the box $i$ and the sum extends over all the boxes; 
$B_1(\beta,q)$ and $B_2(\beta,q)$ are constants,
depending on $\beta$ and $q$, but not on $l$ 
and $\chi(q)$ is the partition function.
From eq.~(\ref{eq1}) it is easy to find the multifractal spectrum for the system:
for $q<\frac{d}{\beta}$, $\alpha = d$ and $f(\alpha) =  d$; for  $q>\frac{d}{\beta}$
$\alpha = d-\beta$ and $f(\alpha) = 0$.
The exponent $\alpha$ describes the scaling of the mass inside a box of size $l$ as 
$l\rightarrow 0$, and $N\propto l^{-f(\alpha)}$ is the number of such boxes. 
Such results reveal a homogeneous ($f(\alpha)=d$) distribution of boxes whose average density 
$\rho(l)\propto l^{\alpha}/l^{d} = l^d/l^d$ is constant and a finite ($f(\alpha)=0$) set of boxes
(in this case only one), whose average density scales as $\rho(l)\propto l^{\beta-d}$.

The multifractal analysis, therefore,
 correctly detects the presence 
of the central singularity and of an analytic distribution 
everywhere else.
However, if we consider a system described by eq.~(\ref{eq1}), but
made of discrete set of points, the identification
of scalings by box counting analysis is no more straightforward.
Since the system is not uniform, the local interparticle
distance $\lambda$ is a function of the distance from the center $r$:
$\lambda(r) = \left(A^{-1}r^\beta\right)^{1/d}$.
It is easy to see that, if one considers  boxes of size $l>l_o=A^{-1/d}$
(where $A$ is the amplitude in eq.~(\ref{eq1})), they are occupied 
on average. If, on the other hand, $l<l_o$, one can define a characteristic
distance $r_o$ from the center by the equation $\lambda(r_o) = l$.
The boxes at distances $r>r_o$ contain on average one or no particles,
while the boxes at $r<r_o$ are on average occupied.
In other words, there is a $l$-dependent scale $r_o$ below which the system
is analogous to the continuum case, and above which the system looks intrinsically
discrete.

A major difference between
a fractal (or a multifractal) and a halo described by 
eq.~(\ref{eq1}) is the fact that, in the fractal, the 
density fluctuations are large at any scale. In the halo, instead,
the density varies smoothly.
A valid candidate  to quantify such fluctuations
 is the conditional variance, defined as the
mean square density fluctuation in spheres centered on points
of the system, normalised to the average conditional density (eq.~(\ref{eq4}))
\cite{GEF}:
\be
\sigma_c^2 (r) = \frac{\langle\rho_R(r)^2\rangle_c - \langle\rho_R(r)\rangle^2_c}
{\langle\rho_R(r)\rangle^2_c},
\ee
where $\rho_R(r)$ is the density in a sphere centered in $R$ with radius
$r$, and the subscript $c$ means that the corresponding quantities
are ``conditional''.
In particular, $\langle \rho_R(r)^2\rangle_c$, where 
the average $\langle...\rangle_c $ is performed over all occupied points,
can actually be rewritten as $\langle \rho_R(r_0)\rho_R(r)\rho_R(r) \rangle $ 
where the average $\langle...\rangle $ is performed 
over all the $\vec{r_0}$ in the volume. In turn,   
$\langle \rho_R(r_0)\rho_R(r)\rho_R(r) \rangle $ is actually the  three point correlation function $\langle \rho_R(r_i)\rho_R(r_j)\rho_R(r_K) \rangle $
with $r_i = r_j$. 
This shows that $\langle \rho_R(r)^2\rangle_c$ is in fact closely related to
the three-point correlation function

In general, for a point distribution, 
$\sigma_c^2(r)$ will be given by the sum of two terms:
$\sigma_c^2(r)= \sigma^2_P(r)+\sigma^2_i(r)$,
where $\sigma^2_P(r)=(\langle\rho_R(r)\rangle_c V(r))^{-1}$ is the variance due to Poissonian noise
 and  $\sigma^2_i(r)$
is the intrinsic variance of the system, which depends on its specific properties.

\begin{figure}[h]
\centerline{
        \psfig{file=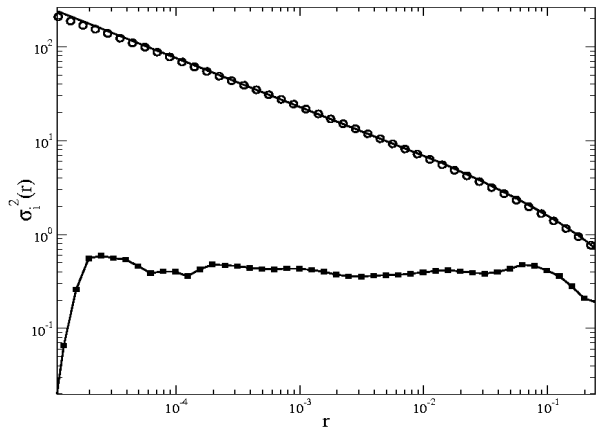,width=8.cm,height=5.8cm}
}
\caption{Normalized conditional variance $\sigma_i^2(r)$ for the samples 
described in fig.~\ref{fig4}. {\em Solid line with squares}: fractal; 
{\em empty circles}: 
halo. 
{\em Solid line}: theoretical result from eq.~(\ref{eq8})
}
\label{fig3b}
\end{figure}

It is possible to compute $\sigma^2_i(r)$ for a cluster
 described by eq.~(\ref{eq1}):
\begin{equation}
\label{eq8}
\sigma^2_i(r)= \frac{1}{d-\beta}\cdot\frac{\left[ \frac{1}{d-3\beta}
\left(1-r^{d-3\beta} \right) + \frac{d^2 r^{d-3\beta}}{(d-\beta)^3} 
\right]}{\left\{ \frac{1}{d-2\beta} + r^{d-2\beta}\left[ 
\frac{d}{(d-\beta)^2}-\frac{1}{d-2\beta}\right]\right\}^2}-1.
\end{equation}

From eq.~(\ref{eq8}) it is easy to see that for $\beta> d/2$, $\sigma^2_i(r)\propto r^{\beta-d}$.
On the contrary, since a fractal is  a scale invariant structure,
 $\sigma^2_i(r)$ (often referred to as 
{\em lacunarity}~\cite{GEF,manlac,BLU}) is constant.
In fig.~\ref{fig3b} we plot $\sigma^2_i(r)$ for a fractal and a halo,
together with the analytic result of eq.~(\ref{eq8}).

In addition to the conditional variance we introduce a new statistical 
concept, the 
 ``scaling gradient'' $\Delta$, which permits also a local analysis
of the fluctuations.

\begin{figure}[h]
\centerline{
        \psfig{file=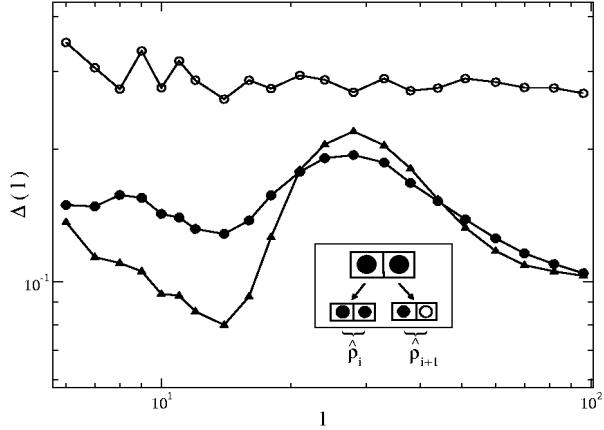,width=8.cm,height=6.cm}
}
\caption{Measure of the scaling gradient $\Delta(l)$ as a function
of the box size $l$ for different samples in 
$3d$. {\em Empty circles}: 
a fractal with dimension  $D=2$ generated by a Levy
flight algorithm. {\em Filled circles}: a halo
with $\beta=2$. {\em Triangles}: a homogenous set. 
In the inset, the procedure followed to measure $\Delta(l)$ in $1d$: two
 adjacent occupied boxes (the boxes with filled circles) are divided
in two offsprings  each. The offsprings  
of the left box are both occupied ($\hat \rho=1$);
while just one of the offsprings of the right one is occupied ($\hat \rho=1/2$).
The resulting $\Delta$ is $1/2$.  
}
\label{fig10}
\end{figure}
Consider a point distribution in 
 $d$ dimensions extending 
over a finite volume.
The volume is divided in $N_{box}$ identical boxes of size $l$,
 with  the number of occupied boxes being $N_{occ}(l)$. 


We identify all the adjacent pairs of occupied boxes $N_i(l)$, 
where $i$ runs over all the occupied {\it adjacent boxes}, $N_{adj\_occ}$.
Each box {\it i} of the occupied ones is divided in $N_s(i)$ identical boxes (offsprings); some of these will be occupied and we denote 
them as $N_{s\_occ}(i)$.
$N_{s\_occ}(i)$ is the number of occupied offsprings in the box {\it i} and 
let us define  $\hat\rho_i=N_{s\_occ}(i)N_{s}(i)^{-1}$ as the fraction
 of occupied offsprings of the box $i$.

The {\em scaling gradient} of the system is defined as:
\be
\Delta(l)= \frac{1}{N_{occ}(l)}\sum_{i=1}^{N_{adj\_occ}}|\hat\rho_i(l)-
\hat\rho_{i+1}(l)|, 
\ee
where the sum extends over all pairs of adjacent occupied boxes 
$N_{adj\_occ} $.
This measure has the following properties:

(i)  it is a conditional measure, since it only considers
occupied adjacent pairs;

(ii)  it  considers the occupation density $\hat\rho$,
which is a measure of how the occupation of the boxes
scales in the system;

(iii)  it is sensitive to  local fluctuations of $\hat\rho$, although it
is averaged over the whole system.

In other words, the scaling gradient measures the fluctuations
of the fragmentation properties of the system. 

The results of a measure of $\Delta(l)$ in three different $3d$ samples 
are shown in fig.~\ref{fig10}.
While the measure of $\Delta(l)$ for
the homogeneous set and the halo shows a
 peak at a characteristic scale, 
the fractal distribution has a flat $\Delta(l)$.

The behavior of $\Delta(l)$ for
the halo can be explained
as follows.
For $l$ such that $r_o(l) =(Al^3)^{1/\beta}>>1$  all boxes and their
``offsprings'' are occupied: in this case, $\Delta(l)\simeq 0$.
When $l$ is such that $r_o(l)\alt 1$, all the boxes are occupied, but 
some of their ``offsprings'' (with distance from
the center $r\simeq 1$) will be empty. Therefore $\Delta(l)$
grows.
Eventually, when $l$ is such   $r_o(l) \simeq 1$, all the boxes
will be occupied on average. Consider now the generation
of box offsprings in this case:
their  size  is such that 
a large number of them is empty. In particular, it is
the  maximum number of empty boxes deriving from occupied boxes.
For this reason, $\Delta(l)$ reaches a maximum. 
This is apparent in fig.~\ref{fig10}.
On the contrary, since a fractal is a scale invariant structure,  
$\Delta(l)$ is constant at all scales larger than the lower cut-off.
The scaling gradient is therefore able to detect
unambiguously the
scaling properties of different systems characterised
by the same two-point correlations.

In summary, N-body simulations provide evidence for the formation of halos,
clusters which are not really fractals, but still are characterized by power law 
correlations. The galaxy distribution, instead, appears more compatible
with the fractal behaviour in a range of scales.
We have addressed the fundamental issue of the discrimination 
between the two distributions in such a way to offer a series of tools 
which permit clarification of this problem.
This requires going beyond the two point correlations, although with a careful
critical analysis.
For example, we show that the multifractal approach is not suitable in this respect.
The conditional variance is more appropriate for the global properties at large scales, but
for the more relevant case of local scaling, we introduce the new concept of 
``scaling gradient''.
These methods and their critical analysis  
will represent a crucial element for extracting the relevant
statistical 
properties in future large galaxy catalogues and N-body simulations.

\acknowledgments
We are grateful to prof. M. A. Munoz for careful reading and stimulating comments
on the manuscript.

\end{document}